\documentclass[aps,prl,twocolumn]{revtex4}
\usepackage{graphicx}
\usepackage{verbatim}

\newcommand{\be}{\begin{equation}}
\newcommand{\ee}{\end{equation}}
\newcommand{\bea}{\begin{eqnarray}}
\newcommand{\eea}{\end{eqnarray}}
\newcommand{\dd}{\textrm{d}}
\newcommand{\pr}[1]{\left(#1\right)}
\newcommand{\cro}[1]{\left[#1\right]}

\newcommand{\BE}{\begin{eqnarray}}
\newcommand{\EE}{\end{eqnarray}}
\newcommand{\BEn}{\begin{eqnarray*}}
\newcommand{\EEn}{\end{eqnarray*}}
\newcommand{\barr}{\begin{array}}
\newcommand{\earr}{\end{array}}

\newcommand{\bit}{\begin{itemize}}      
\newcommand{\eit}{\end{itemize}}
\newcommand{\bc}{\begin{center}}
\newcommand{\ec}{\end{center}}
\newcommand{\ben}{\begin{enumerate}}    
\newcommand{\een}{\end{enumerate}}

\newcommand{\eps}{\epsilon}

\newcommand{\e}{\mbox{e}}

\begin{document}

\title{Optimal approximations of power-laws with exponentials}
\author{Thierry Bochud}
\affiliation{Nestec Ltd, Av. Nestl\'e 55, 1800 Vevey, Switzerland} 
\email{thierry.bochud@nestle.com}
\author{Damien Challet}
\affiliation{Nomura Centre for Quantitative Finance, Mathematical Institute, Oxford University, 24--29 St Giles', Oxford OX1 3LB, United Kingdom}
\email{challet@maths.ox.ac.uk}
\date{\today}


\begin{abstract}
We propose an explicit recursive method to approximate a power-law with a finite sum of
weighted exponentials.  Applications to moving averages with long memory are discussed in relationship with stochastic volatility models.
\end{abstract}

\maketitle
Exponential moving averages are widely used as tool for computing efficiently averages of time-changing quantities such as volatility and price. Their main advantage resides in their recursive definition that allows for easy numerical implementation, or remarkably simple models of stochastic volatility, such as GARCH \cite{GARCH}. Their use is however conceptually questionable when the process in question has long memory, as the volume and volatility do \cite{Daco,BouchaudPotters,StanleyVol}. One should rather consider a power-law kernel; this requires however considerably more computing power as one must keep track of all the data points. Some authors approximate a power-law with a sum of exponentials in the literature, the record being held by Ref. \cite{600exp}, which uses 600 exponentials for 2 decades but notices that only a few have a significant contribution to the final function.

While the principle of economy should dictate to fit power-law-looking data with nothing else than a power-law (see for instance the controversy in the June 2001 issue of Quantitative Finance), computing real-time averages with a power-law kernel is much eased by the use of a sum of exponentials. Recent stochastic volatility models for instance use a sum of exponentials \cite{ZumbachLynch,ZumbachStochVol,BorlandBouchaudStochVol} (5, 12 and an infinity, respectively) with algebraically decreasing weights and algebraically increasing characteristic times, thereby respecting the long-memory of the volatility, which might explain in part their forecasting performance\footnote{Models with long memory (see also \cite{FisherCalvet,BacryMuzy}) appear to be intrisically better for forecasting \cite{LuxVolForecasting}.}. It is clear that only a handful of exponentials are required in order to approximate a power-law {\em up to a given order of magnitude}, as many practitioners are aware (see for instance \cite{Daco,ZumbachStochVol}). Since financial market data time series do not extend over an infinite period, such approximation will be good enough for application to financial time-correlations. How many exponentials should be used and with what parameters seem never discussed in the literature. Here, we aim to derive an explicit and new simple scheme that improves the often used approximation; in addition we show that the usual assumption of independent contribution from each exponential implies the existence of an optimal number of exponentials.

Let $f(x)=x^{-\alpha}$ and $g(x)=\sum_{i=0}^{N} g_i(x)$ where $g_i(x)=w_i\exp(-\lambda_i x)$. Assume that one would like to approximate $f$ with $g$ from $x=1$ to $x=10^{k}$, that is, over $k$ decades. The standard approach (see for instance \cite{Svensson}) consists in defining a cost function per decade that is the integral of some measure of the difference between $f$ and $g$, i.e.
\be
C=\frac{1}{k}\pr{\int_1^{10^{k}}[-\alpha\log x-\log g(x)]^2\dd \log x}^{1/2},
\ee
and to minimize $C$ with respect to $w_i$ and $\lambda_i$, so as to obtain
$2(N+1)$ coupled non-linear equations. Ad-hoc numerical methods have been
investigated a long time ago, that solve the resulting set of equations by using the Gram-Schmidt orthonormalisation of exponentials \cite{Svensson}. Our aim here is to obtain a sub-optimal (with respect to $C$) but explicit set of $w_i$ and $\lambda_i$.

The proposed method relies on a simple ansatz for $w_i$ and $\lambda_i$. Instead of trying to solve an intricate set of non-linear equations, one observes that the nature of a power-law is to be scale-free, whereas an exponential has a well defined scale. Therefore, the role of each exponential is to approximate a given region of the $k$ decades. In particular, one wishes that the $i$-th exponential approximates correctly $f(x)$ at $x_i=\beta^i$ where $\beta>1$ is a constant. This already suggests that $\lambda_i\propto \beta^{-i}$, which is both intuitive and well-known. Then one matches $g$ to $f$ and its first derivative $g'$ to $f'$ at $x_i=\beta^i$. However, once again, this would yield $2(N+1)$ coupled non-linear equations. The key observation is that, provided that $\beta$ is large enough (see below), only $g_i$ contributes significantly to $g$ at $x_i$, i.e. $g(x_i)\simeq g_i(x_i)$. We therefore solve $g_i(x_i)=f(x_i)$ and $g'(x_i)=f'(x_i)$, which gives
\bea
\lambda_i&=&\alpha\beta^{-i}\\w_i&=&\pr{\frac{e}{\beta^i}}^\alpha.
\eea
However, $g(x_i)>f(x_i)$ because the contribution of the exponentials other than the $i$-th cannot be totally ignored. Therefore, one must correct the above over-optimistic assumption by considering that $g$ is a weighted sum of $g_i(x)$
\be
g(x)=\sum_{i=0}^{N} c_i\beta^{-i\alpha} \exp(\alpha) \exp(-\alpha/\beta^i x),
\ee
where $\{c_i\}$ is a set of correction factors. The last step is to solve $g(\beta^j)=f(\beta^j)$, which is a set of $N+1$ linear equations with variables $c_i$. The complexity of the problem has been greatly reduced. One can solve numerically this set of equations. In order to obtain explicit expressions for $c_i$, one has to resort to another approximation.

The simplest ansatz for $c_i$ already gives a high degree of accuracy and is equivalent to the one currently in use in the literature. Taking uniform $c_i=c$ given by $1/c=\sum_{i=0}^{N} \beta^{-i\alpha}\exp(\alpha) \exp(-\alpha/\beta^i
)$ ensures the equality $g(1)=f(1)$. With this choice the factor $\exp(\alpha)$ disappears from $g(x)$ and
\be
g(x)=\big(\sum_{i=0}^{N} \beta^{-i\alpha}\exp(-\alpha/\beta^i)\big)^{-1}\sum_{i=0}^{N} \beta^{-i\alpha} \exp(-\alpha/\beta^i x)
\ee

\begin{figure}
\centerline{\includegraphics*[width=0.4\textwidth]{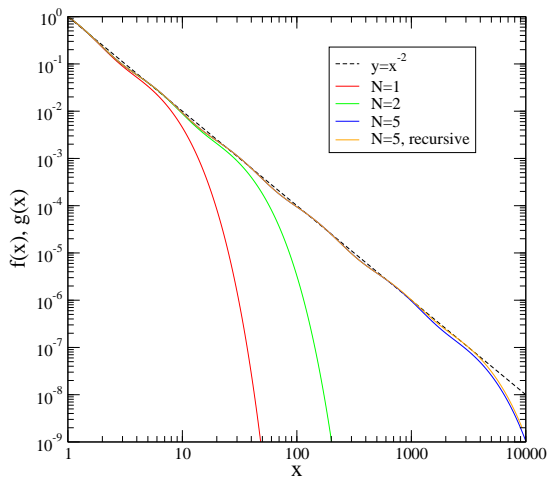}}
\caption{Convergence of the approximation function $g(x)$ to $f(x)$ for the uniform ansatz with 2 (red line), 3 (green line) and 6 exponentials (blue line), and for the recursive ansatz with 6 exponentials (orange line); $\alpha=2$, $\beta=5$}
\label{fig:approx}
\end{figure}

Fig. \ref{fig:approx} shows how the approximation works for increasing $N$: each additional exponential extends the range that is well approximated by a factor $\beta$. The value of $\beta$ was chosen large enough so as to emphasise the oscillations of $g(x)$ at each $\beta^j$. The uniform ansatz implies that while $g(1)=f(1)=1$, $g(\beta^j)>f(\beta^j)$ for $0<j<N$ since the contribution of each $g_k$ is asymmetric with respect to $\beta^j$; when $j=N$, since there are no additional exponentials from $i>j$ to contribute to $g$, $g(\beta^N)<\beta^{-\alpha N}$ (see Fig. \ref{fig:approx}). This problem is of course negligible when a very large number of exponentials is used; however, since our aim is to use as few exponentials as possible it needs to be addressed.

\begin{figure}
\centerline{\includegraphics*[width=0.4\textwidth]{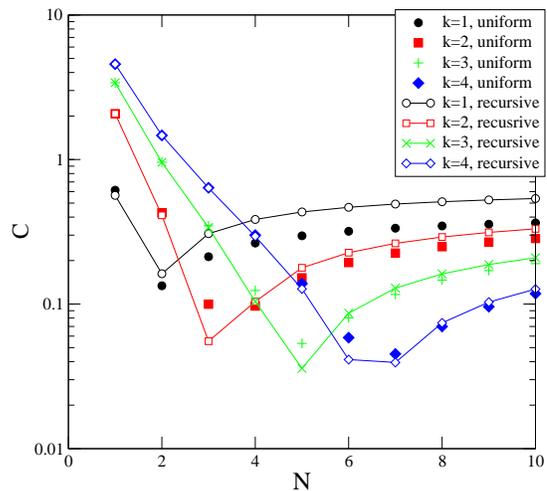}}
\caption{Error per decade $C$ as a function of $N$ for various $k$; $\alpha=2$; $\beta=10^{k/N}$ for the uniform and recursive ansatz (full and empty symbols respectively). Lines are for eye guidance only.}
\label{fig:C_vs_N}
\end{figure}

 The parameter $\beta$ tunes how much of a decade is approximated by a single
 exponential. When $k$ and $N$ are fixed, it is sensible to take $\beta^N=10^k$.
 The cost function $C$ is plotted in Fig. \ref{fig:C_vs_N} as a function of $N$ at fixed $k$ for several values of $k$. For small $N$, $C$ decreases exponentially as a function $N$. Then, strikingly, $C$ has a minimum at $N_m(k)$ and increases slightly before stabilising; the smaller $\alpha$, the smaller the subsequent increase. One would have naively expected that $C$ decreased monotonically as a
 function of $N$; however, since $\beta$ decreases when $N$ increases at fixed
 $k$, the assumption that the exponentials give independent contributions to $g$ is not valid any more at $N\simeq N_m$, and becomes clearly incorrect when
 $N>N_m$. The consequence is that $g(x)$ becomes too large except at $x=1$. This is not problematic, however, since in practice, one prefers large $\beta$ to small ones, so as to use as few exponentials as possible. As expected, $N_m$ increases linearly with $k$, implying that for $\alpha=2$, the optimal $N=N_m(k)\simeq1.7k$, or equivalently $\beta\simeq10^{1/1.7}\simeq3.87$. Another feature of this figure is that $C(N_m(k))$ decreases as function of $k$: this due to the vanishing influence of the deviation caused by the downwards shift of the last exponential.

\begin{figure}
\centerline{\includegraphics*[width=0.4\textwidth]{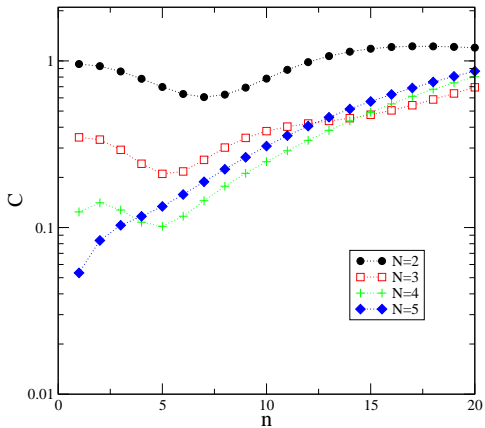}}
\caption{Error per decade $C$ as a function of $n$ for various $N\le N_m=5$; $k=3$, $\alpha=2$; $\beta=10^{k/N}$. Dotted lines are for eye guidance only.}
\label{fig:C_vs_n}
\end{figure}

It is possible to improve the precision of the approximation for $N<N_m$ by modifying the scale of $x$, or equivalently by taking into account derivatives of $g$ of higher orders. The second order yields $\lambda_i=\sqrt{\alpha (\alpha+1)}\beta^{-i}$. From the conditions on the first derivatives and on the equality of functions, $w_i\propto  \beta^{-\alpha i} \exp(\sqrt{\alpha(\alpha+1)})$. This reasoning can be extended to match the derivatives up to order $n$, resulting in
\be
g(x)=\big(\sum_{i=0}^{N} \beta^{-i\alpha}\exp(-\mu/\beta^i)\big)^{-1}\sum_{i=0}^{N}
\beta^{-i\alpha} \exp(-\mu x/\beta^{i} )
\ee
with
\be
mu=\cro{\prod_{j=0}^{n-1} (\alpha+j) }^{\frac{1}{n}}=\cro{\frac{\Gamma(\alpha+n)}{\Gamma(\alpha)}}^{\frac{1}{n}}
\ee
Since $\mu$ does not depend on $i$ it modifies the scale of $x$, which can be
used to adjust the position in log-space of $g$ relative to $f$. For large
$n$, $\mu\simeq (n+\alpha-1)/e$, therefore shifting $g(x)$ to larger
$x$. According to Fig. \ref{fig:C_vs_n}, as long as $N<N_m$, there is an optimal $n$. This comes from the fact that $g(\beta^N)<f(\beta^N)$: it is more advantageous to shift $x$ to larger values so as to avoid the too small value of $g$ at $\beta^N$. It also emphasises once again the need to solve the problem of the last exponential.

The solution comes from a close examination of Fig. \ref{fig:approx}: the first exponentials do not contribute much to the value of $g(\beta^N)$ for $N$ not too small. This suggest that  the contribution of $g_i(\beta^j)$ to $g(\beta^j)$ can be neglected if $i<j$. As a consequence, $g(\beta^N)\simeq g_N(\beta^N)$, and $c_N=1$. Thus 
\be
c_{N-1}=1-\beta^{-\alpha} e^{\alpha(1-1/\beta)}.
\ee
More generally,
\be
c_{N-k}=1-\sum_{i=0}^{k-1}c_{N-i}\beta^{-\alpha(k-i)}e^{\alpha(1-\beta^{i-k})}
\ee

$c_0$ is the same with both ans\"atze, since there is no exponential on the left of $\beta^0$.
Table \ref{table} gives an example set of $c_{N-k}$. It is noticable that $c_{N-k}$ display oscillations which are damped as $k$ increases: since $c_N=1$ is large in order to compensate for the absence of further exponentials, $c_{N-1}$ must be smaller than $c_0$; next, $c_{N-2}$ will be slightly larger than $c_0$ so as to satisfy $g(\beta^{N-1})=f(\beta^{N-1})$, etc.
\begin{center}
\begin{table}[h!b!p!]
\caption{Correction coefficients given by the recursive ansatz. $\alpha=2$, $N=8$, $\beta=4$}
\label{table}
\begin{tabular}{|c|c|c|c|c|c|c|c|c|c|}
\hline $k$&0 &1 &2&3&4&5&6&7&8\\
\hline $c_{N-k}$&1.000 &0.720&0.773&0.763&0.765&0.765&0.765&0.765&0.765\\
\hline
\end{tabular}
\end{table}
\end{center}

\begin{figure}
\centerline{\includegraphics*[width=0.4\textwidth]{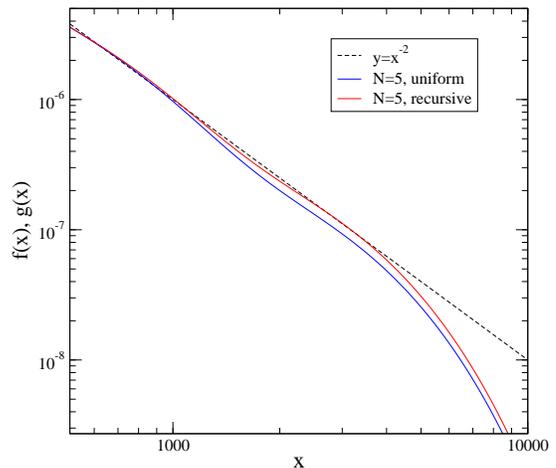}}
\caption{Zoom of Fig \ref{fig:approx} on the last two exponentials. $\alpha=2$; $\beta=4$}
\label{fig:approx_zoom}
\end{figure}

The recursive ansatz always gives a better result that the uniform one, as it ensures that $g(\beta^i)$ is closer to $f(\beta^i)$ for all $i$, and particularly for large $i$; $g$ approximates $f$ remarkably well at $x_i=\beta^i$ provided that $\beta$ is not too small. The differences are most perceptible for $x\simeq \beta^{N}$, where the recursive scheme gives a much better approximation (see Fig. \ref{fig:approx_zoom}), which explains why it is most advantageous for $k\le 4$ where it decreases $C$, at $N_m$ by a factor 2 for $k=2$ and 1.5 for $k=3$; larger $k$, hence larger $N_m$, will not bring much improvement since the weight of the discrepancy caused by the uniform ansatz at $\beta^N$ decreases. Improving the precision further is possible by taking more exponentials from the left hand side of $\beta^j$ into account in the calculus of $c_i$ at the price of heavier and probably non-explicit computations. Finally, if solving the full set of linear equations for $c_i$ does not give enough precision, the remaing possibility is to minimise numerically $C$ \cite{Svensson}.

The above approximation has an obvious application to financial markets.
The measure of historical volatility is usually done with exponential moving averages
\be
V(t+\delta t)=V(t)\Lambda+(1-\Lambda)v(t)
\ee
where $v(t)$ is some measure of the instantaneous volatility (e.g. daily volatility) over $\delta t$ units of time, and $\Lambda=e^{-\lambda}$ is the memory. RiskMetrics recommends $\Lambda_1=0.98$ or $\Lambda_2=0.94$. While this is an efficient way of computing an average, it implicitely assumes a choice of a single time scale $1/|\ln \Lambda|\simeq 1/(1-\Lambda)$ for $\Lambda$ close to 1. Unfortunately, the volatility is a process with no obvious time scale, as its autocorrelation function decreases slowly; fitting it with a power-law gives an exponent  $\nu\simeq0.3$ \cite{BouchaudPotters,Daco}. In other words, any choice of $\Lambda$ is a compromise between smoothness and reactivity. To our knowledge, the first paper to use a power-law kernel for measuring volatilities is from the Olsen group \cite{ZumbachDacoRichter}. One possible reason of this particular functional form of the volatility memory is that the market is made of heteregeneous participants \cite{DacoHeterogeneity}. For instance the variety of time-scales of people taking part into financial markets is obvious to any practioner, hence a choice of a single $\Lambda$ selects the categories of traders that the resulting average volatility incorporates. Direct measure on high-frequency data revealed five characteristic time scales \cite{ZumbachLynch}. Fitting a stochastic volatility model with five time-scales, this work found them to be $0.18$, $1.4$, $2.8$, $7$, $28$ business days, with respective weights of $0.39$, $0.20$, $0.18$, $0.12$, $0.11$; the time scales span about $2.2$ decades, and the weights decreases algebraically as the timescale grows with an exponent of about $\alpha=0.3$. Other work considered $\alpha=2$ \cite{ZumbachDacoRichter,ZumbachStochVol}. Generally speaking, $2\alpha-2=\nu$, which gives $\alpha=1.15$ if $\nu=0.3$ (see e.g. \cite{BorlandBouchaudStochVol}). For $\alpha=1.15$, five exponentials approximate best three decades with corrections $\vec c=(0.704, 0.702, 0.714, 0.647, 1)$. The average volatility $\sigma^2$ is a weighted sum of volatilities on given time scales corresponding to the $\lambda_i$s, which, in principle, still requires to keep the returns over a time horizon equal to the longest time scale; this is barely economical and defeats the initial aim of the approximation. The solution is the use of sums of nested exponential moving averages of the last return that are a proxy for returns on larger time scales \cite{ZumbachLynch,ZumbachOperator}. 

\section{Conclusions}

We have provided a simple method to use efficiently a sum of weighted exponentials as a parsimonious approximation of a power-law with any exponent. In particular, we have shown the existence of an optimal number of exponentials when one neglects the contribution of some exponentials in the determination of the coefficients. The recursive ansatz is probably precise enough for most applications.

We thank Gilles Zumbach for useful discussions.

\bibliography{biblio}

\end{document}